\documentclass[fleqn,usenatbib]{mnras}
\usepackage{newtxtext,newtxmath}

\usepackage[T1]{fontenc}
\usepackage{ae,aecompl}

\usepackage{graphicx}	
\usepackage{amsmath}	
\usepackage{amssymb}	

\usepackage{amsmath}
\usepackage{tikz}
\usepackage{listings}
\usepackage{caption}
\usepackage{animate}
\usepackage{subcaption}
\usetikzlibrary{positioning}
\usepackage[utf8]{inputenc}
\usepackage{setspace}
\usepackage{bm}
\usepackage{caption}
\usepackage{float}
\usepackage{tabularx}
\usepackage{verbatim}
\usepackage{placeins}
\usepackage{color}
\usepackage{ulem}

\newcommand\msun{\,\rm M_\odot}
\newcommand\pc{{\,\rm pc}}

\newcommand\myr{{\,\rm Myr}}

\newcommand\msbh{M_\bullet}

\newcommand\rh{r_{\rm h}}
\newcommand\rmo{r_{\rm m}}
\newcommand\af{a_{\rm f}}
\newcommand\ah{a_{\rm h}}
\newcommand\tnb{T_{\rm NB}}
\newcommand\thdgw{T_{\rm HD}}

\newcommand\se{\sigma_{\rm ecc}}

\title[Black hole coalescence timescales]{Defeating stochasticity: coalescence timescales of massive black holes in galaxy mergers}

\author[I. T. Nasim et al.]{%
Imran Nasim,$^{1}$\thanks{E-mail: i.nasim@surrey.ac.uk (KTS)}
Alessia Gualandris,$^{\!1}$
Justin Read,$^{\!1}$
Walter Dehnen,$^{\!2,3}$ 
Maxime Delorme,$^{\!1,4}$
\newauthor  Fabio Antonini$^{1,5}$
\\
$^{1}$ Department of Physics, University of Surrey, Guildford, GU2 7XH, Surrey, UK\\
$^{2}$ Universit{\"a}ts-Sternwarte M\"unchen, Scheinerstrasse 1, D-81679, Munich, Germany\\
$^{3}$ University of Leicester, Dept.~for Astronomy \& Physics, University Rd, LE1 7RH, UK\\
$^{4}$ D\'epartement d'Astrophysique/AIM, CEA/IRFU, CNRS/INSU, Universit\'e Paris-Saclay, Universit\'e  de Paris, 91191 Gif-sur-Yvette, France,\\
$^{5}$ School of Physics and Astronomy, Cardiff University, Cardiff, CF24 3AA, UK
}

\date{}
\pubyear{2020}

\begin{document}
\label{firstpage}
\pagerange{\pageref{firstpage}--\pageref{lastpage}}
\maketitle

\begin{abstract}
The coalescence of massive black hole binaries (BHBs) in galactic mergers is the primary source of gravitational waves (GWs) at low frequencies. Current estimates of GW detection rates for the {\it Laser Interferometer Space Antenna} and the {\it Pulsar Timing Array} vary by three orders of magnitude. To understand this variation, we simulate the merger of equal-mass, eccentric, galaxy pairs with central massive black holes and shallow inner density cusps. We model the formation and hardening of a central BHB using the Fast Multiple Method as a force solver, which features a $O(N)$ scaling with the number $N$ of particles and obtains results equivalent to direct-summation simulations. At $N \sim 5\times 10^5$, typical for contemporary studies, the eccentricity of the BHBs can vary significantly for different random realisations of the same initial condition, resulting in a substantial variation of the merger timescale. This scatter owes to the stochasticity of stellar encounters with the BHB and decreases with increasing $N$. We estimate that $N \sim 10^7$ within the stellar half-light radius suffices to reduce the scatter in the merger timescale to $\sim 10$\%. Our results suggest that at least some of the uncertainty in low-frequency GW rates owes to insufficient numerical resolution.
\end{abstract}




\section{Introduction}
\label{sec:intro}
Supermassive black holes (hereafter SMBHs) are thought to reside at the centre of most if not all massive galaxies \citep[e.g.][]{kormendyrichstone1995,kormendyho2013} and scaling relations between the SMBH mass and the mass or velocity dispersion of the stellar spheroid suggest they co-evolve with their host galaxy. In the standard cosmological framework of hierarchical structure formation, binaries of SMBHs (hereafter BHBs) form from the merger of two galaxies when each hosts a central SMBH \citep{begelman1980}. While evidence of galaxy interactions is abundant, observations of BHBs have so far revealed only a handful of genuine candidates \citep[e.g.][]{comerford2013}, and for most systems at sub-parsec separations alternative explanations have been put forward \citep[e.g.][]{Heckman1984,Crenshaw2010}. This may suggest that the majority of BHBs harden efficiently and reach coalescence in much less than a Hubble time.

The evolution of BHBs is characterised by three distinct phases \citep{begelman1980}: (i) the dynamical friction phase \citep{chandrasekhar1943}, during which the SMBHs are driven towards the centre of the stellar system by the merging galaxies; (ii) the hardening phase, during which the pair of SMBHs shrinks its separation due to encounters with stars; and (iii) a phase of either stalling or fast inspiral due to the emission of gravitational waves (GWs), depending on whether a significant supply of stars can be provided to interact with the binary. During the hardening phase, stars remove energy and angular momentum from the BHB via the gravitational slingshot mechanism, causing the separation between the SMBHs to shrink \citep{hills1983,quinlan1996}. As stars are removed from the central region, a core is carved in the stellar distribution \citep{milosavljevicmeritt2001}. The subsequent fate of the BHB depends on the supply of stars to the binary's losscone, the region in phase space populated by stars with low enough angular momentum to interact with the BHB. Stalling occurs in spherical systems where two-body relaxation is the only mechanism contributing to losscone refilling, and its characteristic timescale is longer than a Hubble time for all but the smallest galaxies \citep[e.g.][]{1977ApJ...211..244L}. This so-called `Final Parsec Problem' \citep[e.g][]{milo2003} has cast doubt on the likelihood of low frequency GW detections with appreciable rates. However, simulations of galaxy mergers where a BHB is followed from early times show efficient losscone refilling and hardening, leading to BHB coalescence in less than a Hubble time \citep{preto2011, khan2011, GM2012}. This is because the triaxiality of the merger remnant drives angular momentum diffusion in a non-spherical potential, feeding stars into the BHB's losscone \citep{vasiliev2015, G17, bortolas2018}.

The detection of GWs from the coalescence of stellar mass black hole and neutron star binaries \citep[e.g.][]{ligofirst,neutronmerger2017} has marked the birth of GW astronomy, providing unique information on their masses, spins and merger rates. 
Detection of low-frequency GWs from BHBs, the loudest GW sources in the Universe, from missions such as the Laser Interferometer Space Antenna \citep[LISA,][]{LISA_AMARO_SEOANE2017} and the Pulsar Timing Array \citep[PTA, e.g.][]{nanogravpta2015} will constrain the physics of SMBHs, the formation and evolution of BHBs, and the SMBH-galaxy connection.
Current estimates of detection rates for these missions vary widely, with differences of up to three orders of magnitude reported in the literature \citep{wyitheloeb2003,sesana2010}.
In this context, determining merger timescales of BHBs from numerical simulations has become of the utmost importance. Modelling the evolution of BHBs from the kpc-scale of the galaxy merger to mpc-scale of the onset of GW emission is computationally challenging. 
Direct summation codes like \textsc{$\phi$-grape} \citep{harfst2007phigrape} have been successful at accurately modelling binary hardening beyond the hard-binary separation, contributing to the resolution of the Final Parsec Problem. In combination with semi-analytic models of BHB evolution under the combined effects of dynamical hardening and GW emission, they provide estimates of merger timescales varying from tens of Myr to a few Gyr. Due to the $O(N^2)$ scaling imposed by the computation of all pairwise gravitational forces, direct summation methods are limited to about one million star particles, even with the aid of hardware acceleration \citep{vasilievantoninimerritt2014, G17, khantimescalepaper2018}. Relaxation effects are over-represented at such artificially low particle numbers and $N \gtrsim 10^7$ is required for hardening rates to become independent of $N$, a signature that collisionless losscone refilling is at work. \citet{vasiliev2015} adopt a Monte Carlo method in which collisional relaxation can be removed to reach an effective $N\sim 10^8$. However, the technique requires calibration against a direct summation integration and is limited to single-galaxy models. \citet{rantala2016ketju} adopt an extension of the tree/SPH code \textsc{gadget-3} \citep{springel2005gadget} to include chain regularisation in a small region around the SMBHs. This approach can be used to model the evolution of BHBs self-consistently from early times to coalescence \citep{mannerkoski2019}. However, we caution that the large force errors introduced by \textsc{gadget-3} outside the chain can lead to artificially fast hardening, while the prohibitive $O(N^3)$ scaling of the chain limits its applicability to $N\lesssim 50-100$. The Fast Multiple Method (FMM) code \textsc{griffin} is designed to monitor force errors and adaptively select parameters to ensure a distribution of force errors similar to that in a direct summation code while retaining the $O(N)$ scaling of the FMM technique \citep{dehnen2014griffin}. Simulations of isolated triaxial models show that angular momentum diffusion is correctly captured in \textsc{griffin} \citep{G17}. 

Here, we present a modified version of the  \textsc{griffin} code in which SMBH-star encounters are modelled via direct summation, with indistinguishable results from the \textsc{$\phi$-grape} code. We present a set of integrations of equal mass galaxy mergers with shallow inner density profiles and large orbital eccentricity. We follow the evolution of BHBs from early times past the hard-binary separation phase, and extrapolate the evolution of the orbital elements due to dynamical effects and GW emission to estimate coalescence timescales for LISA and PTA sources. Interestingly, we find evidence of large stochasticity in the eccentricity of the BHBs at early times, which translates into large errors in the estimated coalescence timescales. We show that such stochasticity owes to insufficient numerical resolution and we calculate resolution requirements to obtain accurate estimates of merger timescales.

\section{Numerical setup}
\label{sec:setup}
We perform four suites of $N$-body simulations with one suite at a poor resolution (PR) of $N=128\rm k$, one suite at a low resolution (LR) of $N=256\rm k$, two suites at a medium resolution (MR) of $N=512\rm k$ particles and one suite at a higher resolution (HR) of $N=2048\rm k$ particles. We model mergers of equal mass galaxies hosting a central SMBH. Each galaxy follows a \cite{dehnen1993} density profile  representative of a nuclear bulge
\begin{equation}
\label{eqn:dehnen_density}
	\rho(r) = \frac{(3-\gamma)M}{4\pi} \frac{r_0}{r^{\gamma} (r+r_0)^{4-\gamma}}
\end{equation} 
with total mass $M$, scale radius $r_0$ and inner slope $\gamma$. Units are chosen such that $M_{\rm tot}=G=r_0=1$, where $M_{\rm tot}$ is the total stellar mass in the merger. Each galaxy has a shallow $\gamma=0.5$ profile and the SMBH mass is $M_{\bullet} = 0.005$. The star to SMBH mass ratio is approximately $3.2\times10^{-3}$, $1.6\times10^{-3}$, $8\times10^{-4}$ and $2\times 10^{-4}$ for increasing $N$. The two galaxies are placed at an initial distance $R=20r_0$ on a bound elliptical orbit with eccentricity $e=0.9$. A large eccentricity is chosen to mimic merger conditions in cosmological simulations \citep[e.g.][]{khochfar2006eccen} as well as to reduce computational time. The simulation parameters are given in Table~\ref{tab:init}.
\begin{table}
\begin{tabular}{cccccccc}
\hline 
suite & $M_1 : M_2$ & $M_{\bullet}/M_{*}$ & $N$  & $\gamma$ & $R/r_0$ & $e_i$ &  $N_r$ \\
\hline 
PR & 1:1 & 0.005 & 128k & 0.5 & 20 & 0.9 & 16\\
LR & 1:1 & 0.005 & 256k & 0.5 & 20 & 0.9 & 12\\
MR & 1:1 & 0.005 & 512k & 0.5 & 20 & 0.9 & 8\\
HR & 1:1 & 0.005 & 2048k & 0.5  & 20 & 0.9 & 4\\
\hline
\hline
\end{tabular}
\caption{Initial parameters of the galaxy mergers. From left to right: Simulations suite: poor resolution (PR), low resolution (LR), medium resolution (MR) and high resolution (HR); mass ratio between the galaxies; SMBH to stars mass ratio; total number of particles in the merger $N$; inner slope of the galaxy density profile $\gamma$; initial distance between the centres of the two galaxies $R$; initial orbital eccentricity of the progenitor galaxies $e_i$; number of random realisations $N_r$.} 
\label{tab:init}
\end{table}
To investigate the effects of stochasticity we generate sixteen random realisations for the PR suite, twelve for the LR suite, eight for the MR suite and four for the HR suite.

We evolve the PR, LR, MR and HR models with \textsc{griffin} \citep{dehnen2014griffin,G17}, which uses FMM as force solver for star-star gravity, avoiding a tail of large force errors, with mean relative force error of $3\times10^{-4}$ (default \textsc{griffin} setting). SMBH gravity is computed by direct summation and all trajectories are integrated using the leapfrog integrator. To validate this approach we also evolve the MR models with \textsc{$\phi$-grape} \citep{harfst2007phigrape}, a direct summation fourth order Hermite predictor-corrector scheme adapted to run on GPUs via the \textsc{sapporo} library \citep{gaburov2009}.
 
For the \textsc{griffin} simulations we adopt a softening length of $\epsilon_* = 2.3 \times 10^{-2}$ for the stars and $\epsilon_{\bullet}= \epsilon_*/100 = 2.3 \times 10^{-4}$ for the black holes. For the \textsc{$\phi$-grape} simulations instead we used $\epsilon = 10^{-4}$ for all particles which is commonly used in studying the evolution of BHBs \citep[e.g.][]{GM2012}. 

\section{Results}
\label{sec:results}

\subsection{Black hole binary evolution}
\begin{figure}
    \centering
    \includegraphics[width=\columnwidth]{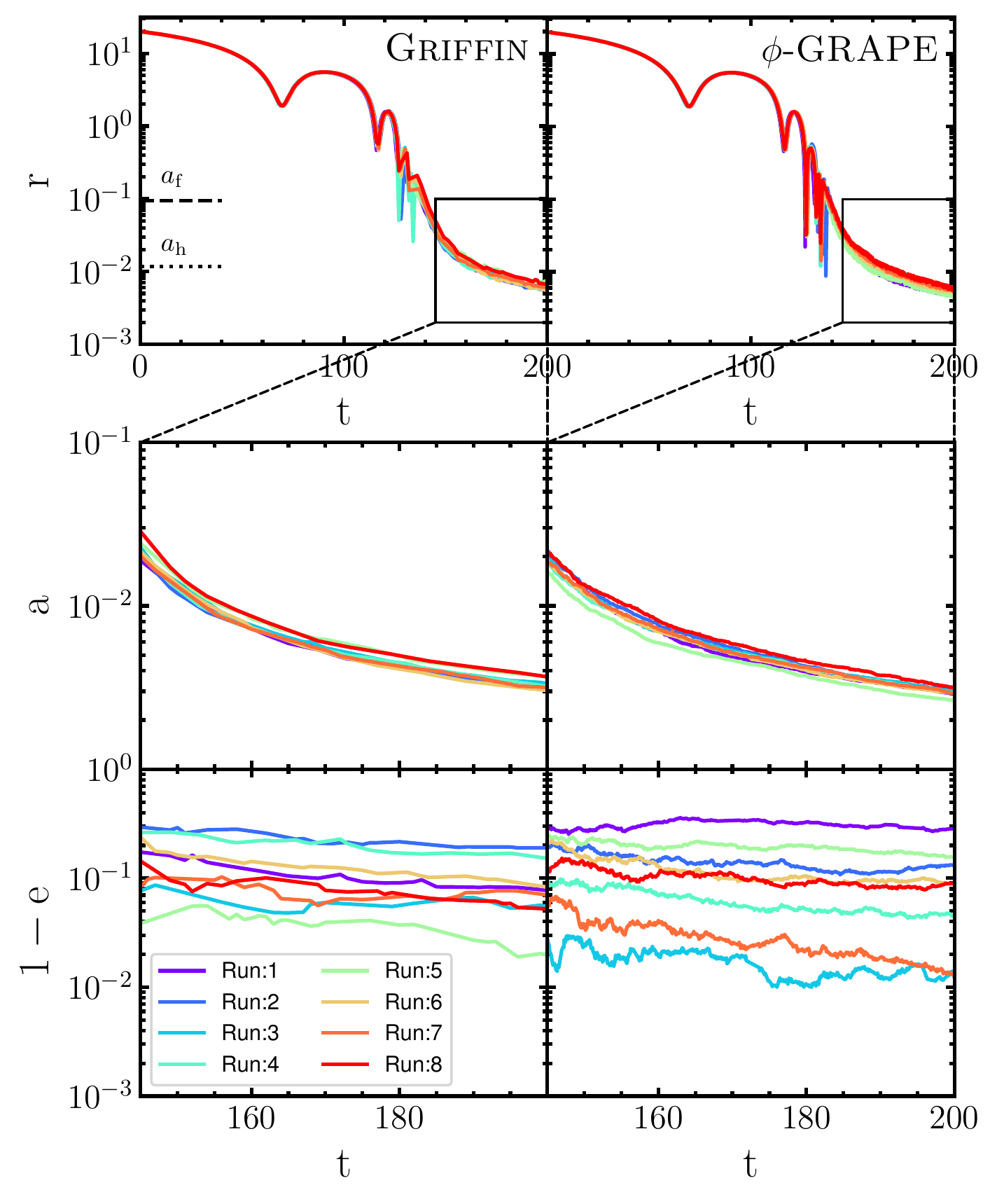}
    \caption{Evolution of the BHB orbital elements as a function of time for the MR suite: distance between the SMBHs (upper panel), semi-major axis (middle panel) and eccentricity (lower panel) in scalable $N$-body units, with the  \textsc{griffin} (left panels) and \textsc{$\phi$-grape} codes (right panels). The relevant separations $\af$, roughly corresponding to the end of the dynamical friction phase, and the hard-binary separation $\ah$, are marked in the top panels.}
    \label{fig:comparison_figure}
\end{figure}
We compare the evolution of the BHB in the \textsc{griffin} and \textsc{$\phi$-grape} integrations to validate the reliability of the binary treatment in the FMM code. All models show the characteristic three phases of binary evolution \citep[e.g.][]{GM2012,bortolas2016}, see top panel in Fig.~\ref{fig:comparison_figure}). In the first phase, the galaxies inspiral and merge due to dynamical friction, bringing the SMBHs to a separation $\af$, defined as the separation at which the stellar mass $M_*$ within the binary orbit is twice the mass of the secondary black hole $M_{\bullet}$:
\begin{equation}
    M_*(\af) = 2M_{\bullet}
\label{eq: ah}
\end{equation}
Around the time the SMBHs reach $\af$ three-body encounters start to become important and these eventually dominate the binary evolution in the second phase of binary hardening. During this rapid phase of strong scatterings, the SMBHs form a bound Keplerian binary and the classical orbital elements can be computed. Stellar ejections lead to a drop in the central density and the formation of a central core. The binary reaches the {\it hard binary separation} $\ah$ when its binding energy per unit mass exceeds the kinetic energy per unit mass of the stars \citep{merritbook2013}
\begin{equation}
\ah = \frac{G\mu}{4\sigma^2} 
\end{equation}
where $\mu$ is the reduced BHB mass and $\sigma$ is the stellar velocity dispersion. For an equal mass binary this reduces to $\ah = GM_{\bullet}/(8\sigma^2)$. An alternative definition that is better suited to $N$-body simulations is given by 
\begin{equation}
    \ah = \frac{\mu}{M_{\rm bin}}\frac{\rmo}{4} = \frac{q}{(1+q)^2} \frac{\rmo}{4}
\label{eq: hard_binary_sep}
\end{equation}
where $M_{\rm bin}$ is the mass of the BHB, $q$ is the black hole mass ratio and $\rmo$ represents the radius containing a mass in stars equal to twice the mass of the primary.
Values of $\af$ and $\ah$ for our models are marked in the top panels of Fig.~\ref{fig:comparison_figure}.

The second phase ends when all stars initially in the binary's losscone have been ejected, and any further hardening depends on the rate of losscone refilling. The only mechanism contributing to scattering stars into the losscone in spherical systems is two-body relaxation. Because the relaxation timescale is longer than a Hubble time in galaxies, this process is inefficient and leads to stalling in the BHB's evolution. A collisionless mode of losscone refilling, however, is available in non-spherical systems, such as merger remnants, leading to sustained hardening down to separations where decay due to emission of gravitational waves becomes dominant.

\begin{figure}
    \centering
    \includegraphics[width=0.8\columnwidth]{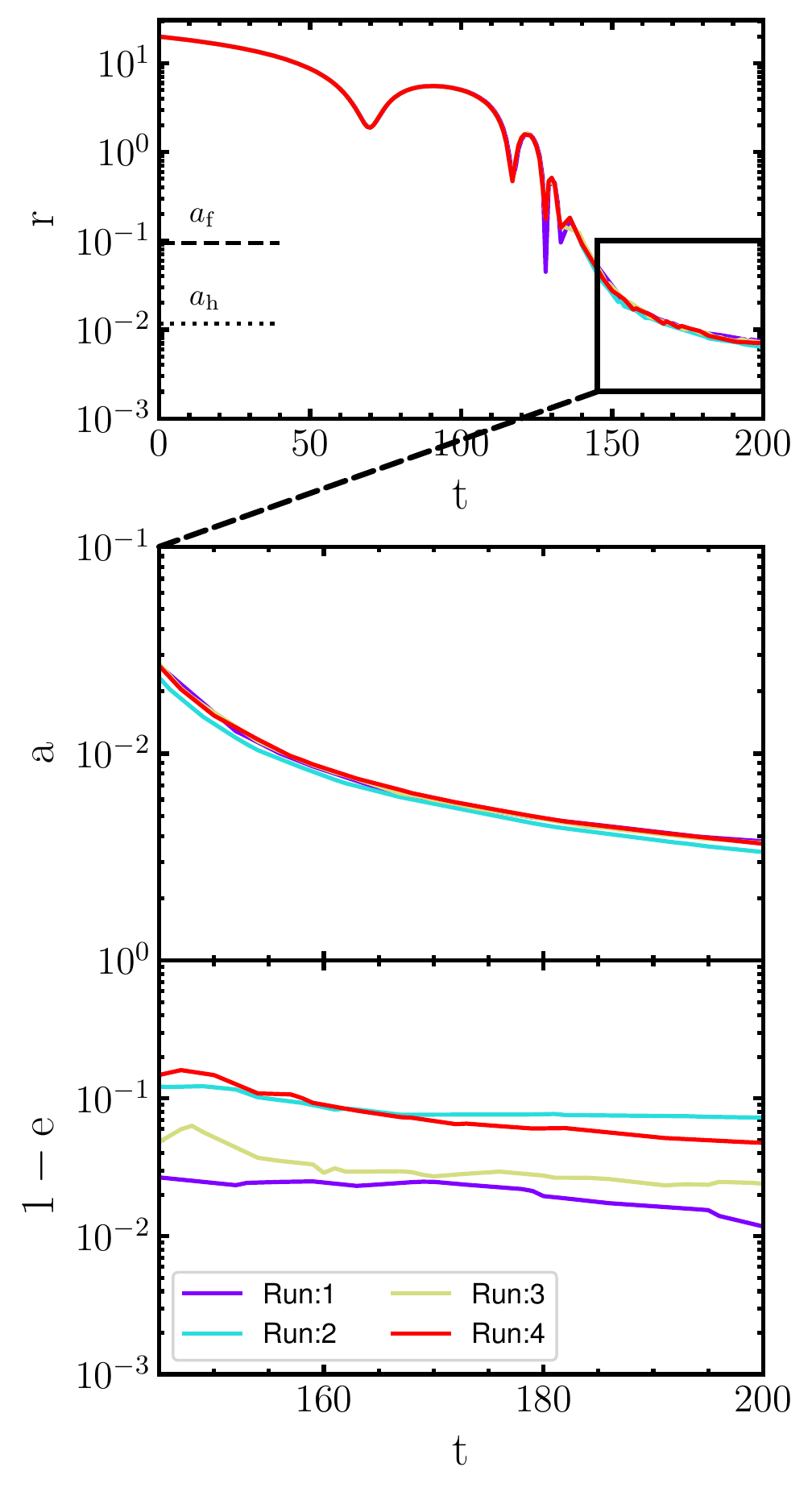}
    \caption{Evolution of the orbital elements of the BHB in the HR simulations. Panels and conventions as in Fig.\-\ref{fig:comparison_figure}.
    }
    \label{fig:high_res_figure}
\end{figure}

The evolution of the binary's orbital elements and separation between the SMBHs is shown in Fig.~\ref{fig:comparison_figure} for both codes. We find that the large scale trajectories of the SMBHs agree remarkably well, as does their relative separation. The semi-major axis and eccentricity evolution are also fully consistent, with direct summation giving a slightly faster decay due to its smaller adopted softening \citep{G17}. We find, however, strong evidence of stochasticity in the eccentricity at the time the BHB becomes bound with a spread in $\log(1-e)$ of about $-0.64$, and a dispersion of about $-1.1$. This can be attributed to  stochasticity in the stellar encounters experienced by the BHB, that determine energy and angular momentum exchanges with the stars. In the case of unrealistically large star to BHB mass ratios, as is inevitably the case in $N$-body simulations, these encounters are responsible for Brownian motion of the BHB \citep{bortolas2016}.
We expect, therefore, an $N$-dependence in the observed spread with an $1/\sqrt{N}$ scaling, as the star to SMBH mass ratio decreases. We find this is confirmed by the HR runs, whose evolution of binary elements is shown in Fig.~\ref{fig:high_res_figure}.
While the evolution in the semi-major axis is practically indistinguishable in the MR and HR runs, the scatter in eccentricity is considerably reduced, 
with a spread in $\log(1-e)$ of about $-1.05$ and a dispersion of roughly $-1.40$.

We note that the binaries in the HR models are characterised by large eccentricities at formation $e\sim 0.9$,  which may suggest a correlation with the initial orbital eccentricity of the merger.

\subsection{Merger timescales}
A spread in the eccentricity at binary formation ought to have a significant effect on the merger timescale of the binaries. To quantify this effect we first extrapolate the evolution of the orbital elements from the end of the $N$-body integrations to coalescence due to emission of GWs. This requires scaling the simulations to physical units. We consider five mass scalings for the SMBH mass ranging from $M_{\bullet}= 10^6\msun$ to $M_{\bullet}= 10^9\msun$, and including the Milky Way black hole (see Table~\ref{tab:scaling}). These are meant to represent both LISA and PTA targets. The host galaxy mass is then set naturally by our assumed galaxy-to-SMBH mass ratio (see Table~\ref{tab:init}), ranging from $M_{*}=2\times{10^8}\msun$ to $M_{*}=2\times{10^{11}}\msun$. Length units are set to match the influence radius of the SMBH. In the case of the Milky Way, for $\rh=3\pc$, this gives $[L] = 30\pc$. For larger SMBH masses, the influence radius is computed assuming a velocity dispersion from the $\msbh-\sigma$ relation \citep{ferrarese2005}. The corresponding time and velocity units are given by $[T] = \sqrt{[L]^3/G[M]}$ and $[V] = \sqrt{G[M]/[L]}$. 

\begin{table}
\centering
\caption{Physical scalings for the $N$-body simulations: SMBH mass, radius of influence, hard binary separation, length scale, and time scale.}
\begin{tabular}{cccccc}
\hline 
Scaling & $M_{\bullet}$ & $\rh$ & $\ah$ & [$L$] & [$T$]\\
 &($\msun$) & ($\pc$) & ($\pc$) & ($\pc$) & ($\myr$)  \\
\hline 
A  &  $10^6$ & $0.95$ & $0.12$ & $10$ & $2.36\times 10^{-2}$ \\
B  &  $4\times 10^6$ &  $2.9$ & $0.35$ & $30$ & $6.12\times 10^{-2}$ \\
C  &  $10^7$ & $3.63$ & $0.45$ & $38$ & $5.52\times 10^{-2}$ \\
D  &  $10^8$ & $13.3$ & $1.65$ & $140$ & $1.23\times 10^{-1}$ \\
E  &  $10^9$ & $52.4$ & $6.49$ & $550$ & $3.04\times 10^{-1}$ \\
\hline
\end{tabular}
\label{tab:scaling}
\end{table}

\begin{figure}
\centering
\includegraphics[width=\columnwidth]{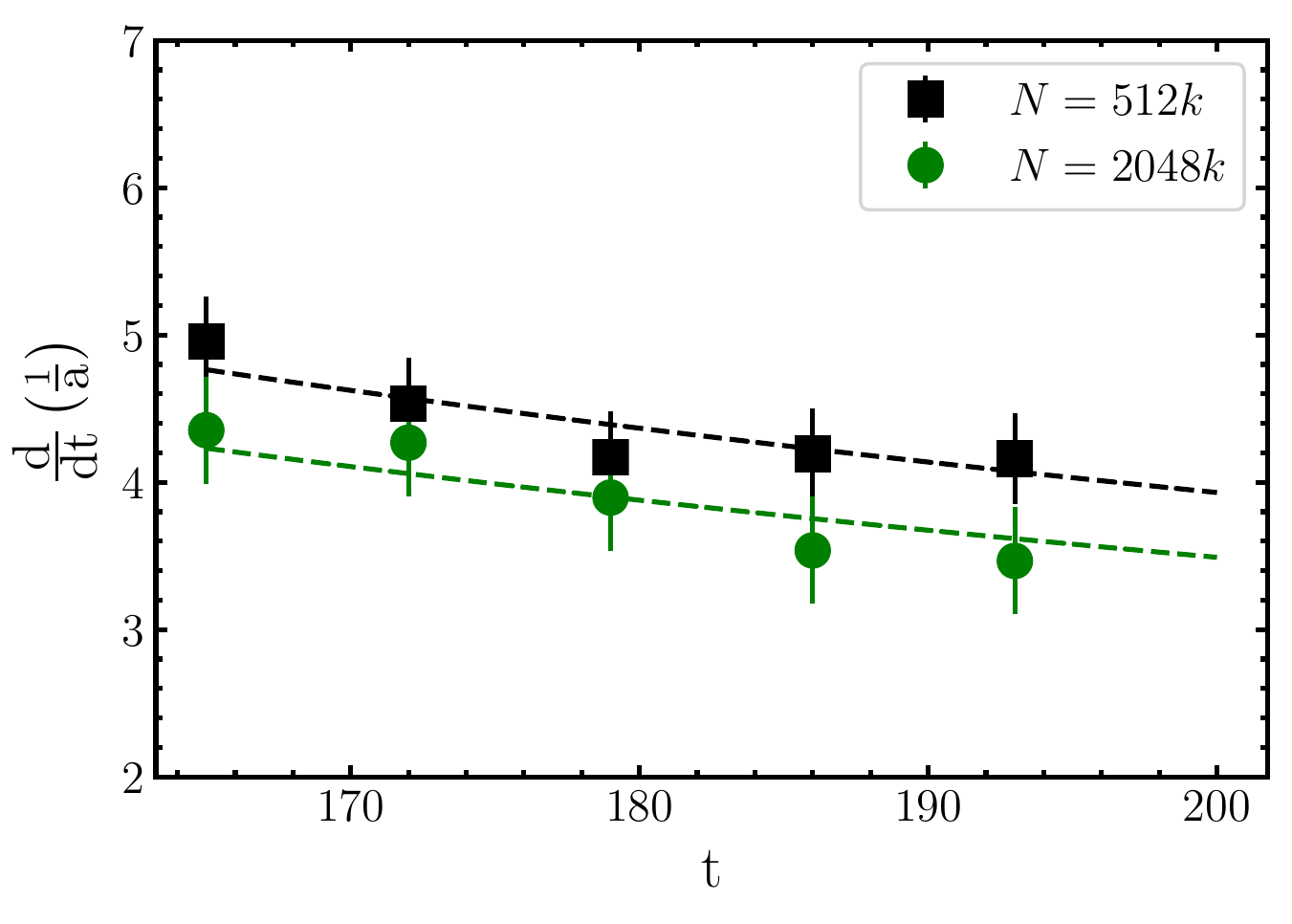}
\caption{Hardening rate (as defined in Eq.~\eqref{eq:s}) as a function of time, averaged over the different random realisations for the MR  (squares) and the HR  (circles) \textsc{griffin} runs, with 1$\sigma$ error bars. The hardening rate predicted by the semi-analytical model is shown by the dashed lines.}  
\label{fig:hard_rate}
\end{figure}

We compute the time-dependent hardening rate of the BHB in all models as \citep{quinlan1996}
\begin{equation}
s = \frac{d}{dt} \left(\frac{1}{a} \right)
    \label{eq:s}
\end{equation}
from the time of binary formation to the end of the $N$-body integrations. This is shown in Fig.~\ref{fig:hard_rate} for both the MR and HR runs.
As seen in previous works \citep{vasiliev2015, bortolas2016}, the hardening rate slowly decreases over time due to the losscone region becoming smaller as the binary shrinks. We also observe a small $N$-dependence consistent with residual collisional effects at these resolutions \citep{G17}.

In order to take into account the slow decline of $s$, we fit the time-dependent hardening rate computed from the $N$-body simulations with a polynomial extrapolation. In this method, called the \textit{Continuous Coefficients Method} (CCM), the hardening rate takes the form 
\begin{equation}
s(t) = \sum_{i=1}^{N} A_{i}\left(\frac{t_{0}}{t}\right)^{i}   
\label{eq:hard_rate_model}
\end{equation}
where $A_i$ are the numerical coefficients of order $i$. In this polynomial extrapolation each subsequent order is inclusive of the previous order but with the addition of a discrete term. The expansion orders can be written as
\begin{subequations}
	\begin{align}
	&S_{1} = A_{1} \left(\frac{t_0}{t}\right)\\
	&S_{2} = S_{1} + A_{2}\left(\frac{t_0}{t}\right)^2
	\end{align}
\end{subequations}
where $t_0$ is the time at which the binary reaches the hard binary separation. 
The numerical coefficients $A_{i}$ are determined by fitting the functional form of equation \eqref{eq:hard_rate_model} to the hardening rate data.
This model-independent extrapolation approach benefits from a functional form that is essentially a perturbed linear extrapolation but with the addition of higher order terms, ensuring a faster convergence. In addition, because no constant term is present, the model can fit arbitrarily small hardening rates, typical, for example, of stalled binaries in the context of the  ``final parsec problem". For the extrapolations in this study we consider a first order expansion. Predictions for the MR and HR runs are shown by the dashed lines in Fig.~\ref{fig:hard_rate}. They all fall well within the one $\sigma$ error bars of the numerical hardening rates.

To estimate the merger timescales of the binaries, we adopt a semi-analytic model of the binary evolution past the end of the numerical integrations that incorporates the effects of both three-body encounters with stars, and GW emission
\begin{subequations}
\label{eq:peters_plus_hard_rate}
\begin{align}
&\frac{da}{dt} =  \frac{da}{dt}\bigg|_{\rm 3B} +
\frac{da}{dt}\bigg|_{\rm GW} = -s(t)a^2(t) + \frac{da}{dt}\bigg|_{\rm GW}\\
&\frac{de}{dt} = \frac{de}{dt}\bigg|_{\rm GW},
\end{align}
\end{subequations}
where $s(t)$ is the time-dependent hardening rate predicted by our CCM extrapolation. For the GW term we adopt the prescription by \citet{peters1964} 
\begin{subequations}
\label{eq:peters64}
\begin{align}
&\frac{da}{dt}\bigg|_{\rm GW} = -\frac{64}{5} \beta \frac{F(e)}{a^3}\\
&\frac{de}{dt}\bigg|_{\rm GW} = -\frac{304}{15}\beta \frac{e G(e)}{a^4},
\end{align}
\end{subequations}
where
\begin{subequations}
\label{eq:peters64_params}
\begin{align}
&F(e) = \left(1-e^2\right)^{-7/2} \left(1 + \frac{73}{24}e^2 +
\frac{37}{96} e^4\right), \\
&G(e) = \left(1-e^2\right)^{-5/2} \left(1 + \frac{121}{304}e^2 \right), \\
&\beta = \frac{G^3}{c^5} M_1 M_2 \left(M_1+M_2\right). 
\end{align}
\end{subequations}

\begin{figure}
    \centering
    \includegraphics[width=\columnwidth]{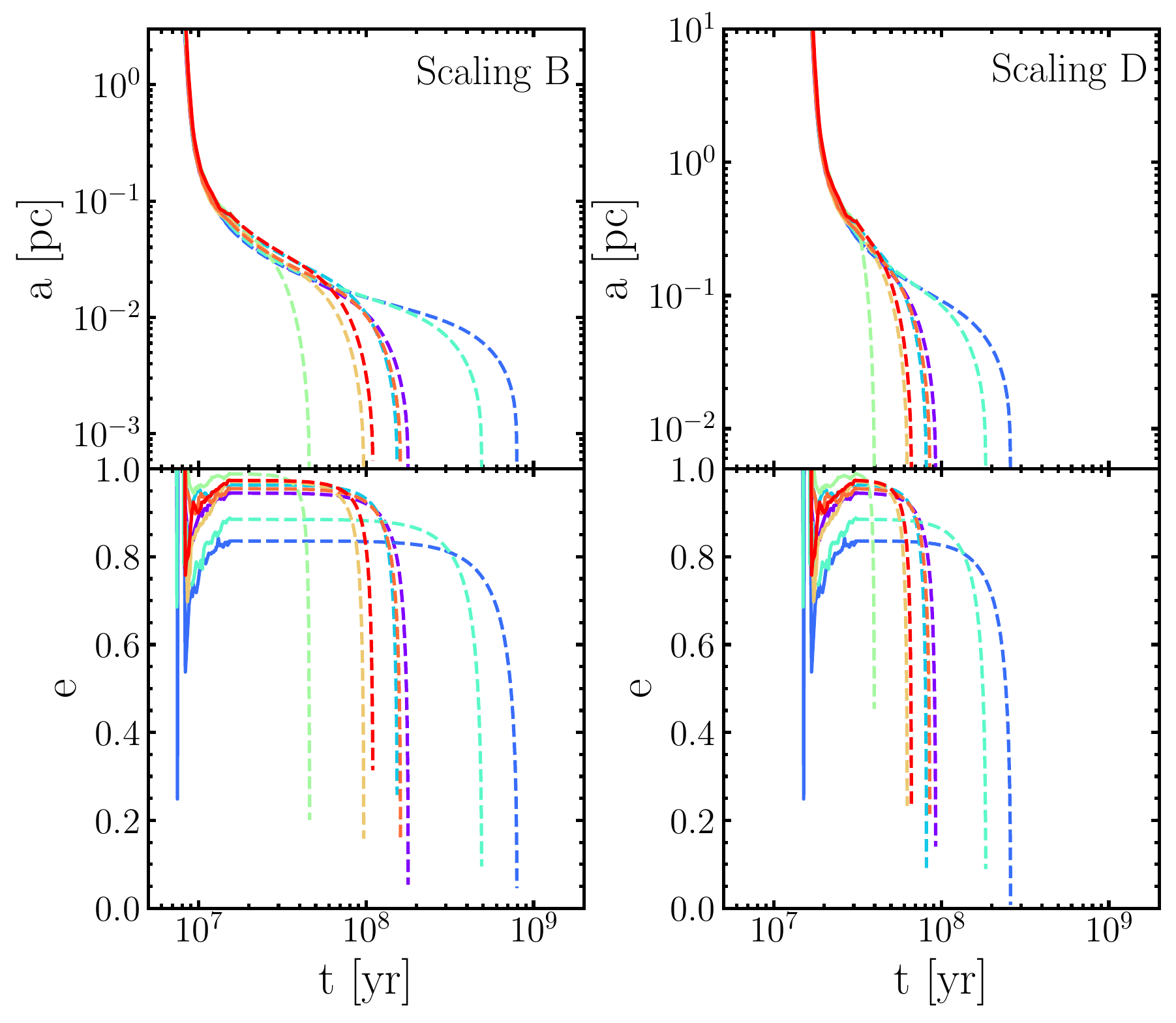}
    \caption{Evolution of the orbital elements for the BHBs under the assumption of a Milky Way type scaling (scaling B) and a larger BH mass scaling (scaling D) relevant for PTA missions. The $N$-body evolution is given by the solid lines and the extrapolated evolution obtained using the CCM method is given by the dashed lines. Different colours refer to the different random realisations of the MR models, as in Fig.~\ref{fig:comparison_figure}.}
    \label{fig:extrap_figure}
\end{figure}
The resulting evolution of the orbital elements from binary formation to coalescence is shown in Fig.~\ref{fig:extrap_figure} for the MR models. Merger timescales vary by as much as an order of magnitude in these models due to the scatter in eccentricity at binary formation. Replacing our CCM extrapolation method with a constant hardening extrapolation results in a similarly large variation in merger times, showing that this is not due to the specific choice of hardening model.

\begin{figure}
    \centering
    \includegraphics[width=\columnwidth]{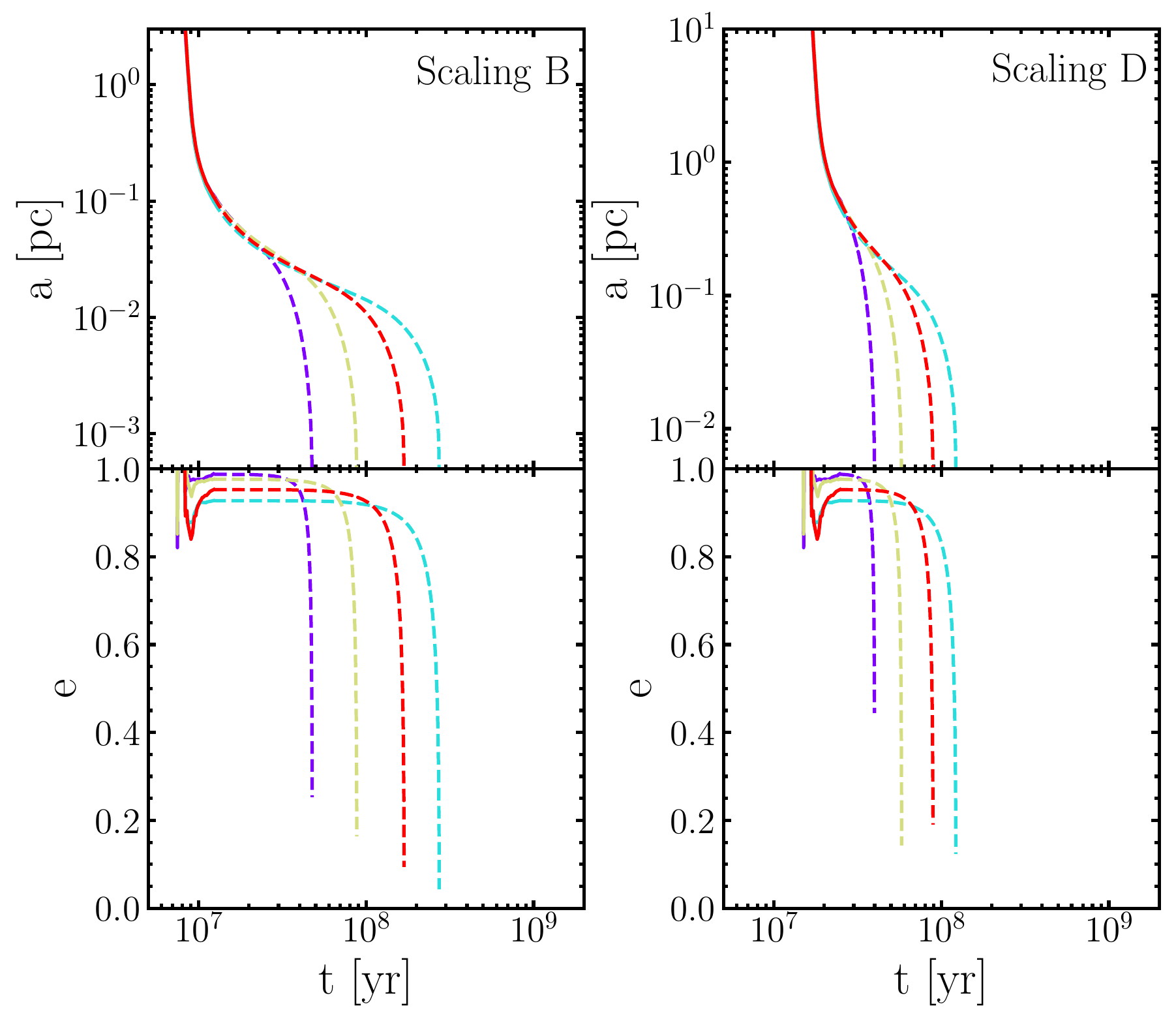}
    \caption{The same as Fig.~\ref{fig:extrap_figure} but for the HR models.}
    \label{fig:high_res_extrap}
\end{figure}
A significant reduction in the spread of merger timescales is observed in the HR models (see Fig.~\ref{fig:high_res_extrap}), where times differ by at most a factor of three.

\begin{table}
\begin{center}
\caption{Binary merger phase parameters. From left to right: Scaling, total particle number in the simulation, $N$-body integration time in \textsc{griffin}, mean extrapolated time to coalescence due to both dynamical hardening and GW emission, dispersion of the merger timescale, dispersion of the eccentricity at binary formation, mean eccentricity at the end of the $N$-body integration.}
\label{tab:merger_time_params}
\begin{tabular}{c c c c c c c}  
\hline 
Scaling & $N_{T}$ & $\tnb$ & $\langle\thdgw\rangle$ & $\sigma_{\rm m}$ & $\sigma_{\rm ecc}$ & $\langle e_{\rm final} \rangle$ \\
      &  & [Myr]  & [Myr] & [dex] & &  \\
\hline 
A  &  $512k$ &  $4.71$ & $130.65$ &  $8.07$ & $0.068$ & $0.94$ \\
   &  $2048k$ &  $4.71$ & $69.24$ & $7.55$ & $0.032$ & $0.96$ \\  
B  &  $512k$ & $12.25$ & $241.34$ &  $8.32$ & $0.068$ & $0.94$ \\
   &  $2048k$ & $12.25$ & $132.04$ & $7.82$ & $0.032$ & $0.96$ \\
C  &  $512k$ & $11.04$ & $102.4$ &  $7.91$ & $0.068$ & $0.94$ \\
   &  $2048k$ & $11.04$ & $59.8$ & $7.46$ & $0.032$ & $0.96$ \\
D  &  $512k$ & $24.70$ & $84.16$ &  $7.78$ & $0.068$ & $0.94$ \\
   &  $2048k$ & $24.70$ & $52.57$ & $7.34$ & $0.032$ & $0.96$ \\
E  &  $512k$ & $60.81$ & $87.07$ &  $7.78$ & $0.068$ & $0.94$ \\
   &  $2048k$ & $60.81$ & $54.73$ & $7.43$ & $0.032$ & $0.96$ \\
\hline
\end{tabular} 
\end{center}
\end{table}
Coalescence times are given in Table~\ref{tab:merger_time_params}, split into the time spent in the $N$-body integration and time spent from the end of the integration to coalescence. The total time from the onset of the galactic merger to coalescence is given by the sum of these two times. We find a marked difference in the coalescence time of the MR and HR models, as well as a clear trend for a smaller dispersion in the merger times at higher resolution. Given the strong dependence of the GW timescale on eccentricity, models with a higher eccentricity at binary formation reach the GW phase earlier than models with a lower eccentricity. 
\begin{figure}
    \centering
    \includegraphics[width=\columnwidth]{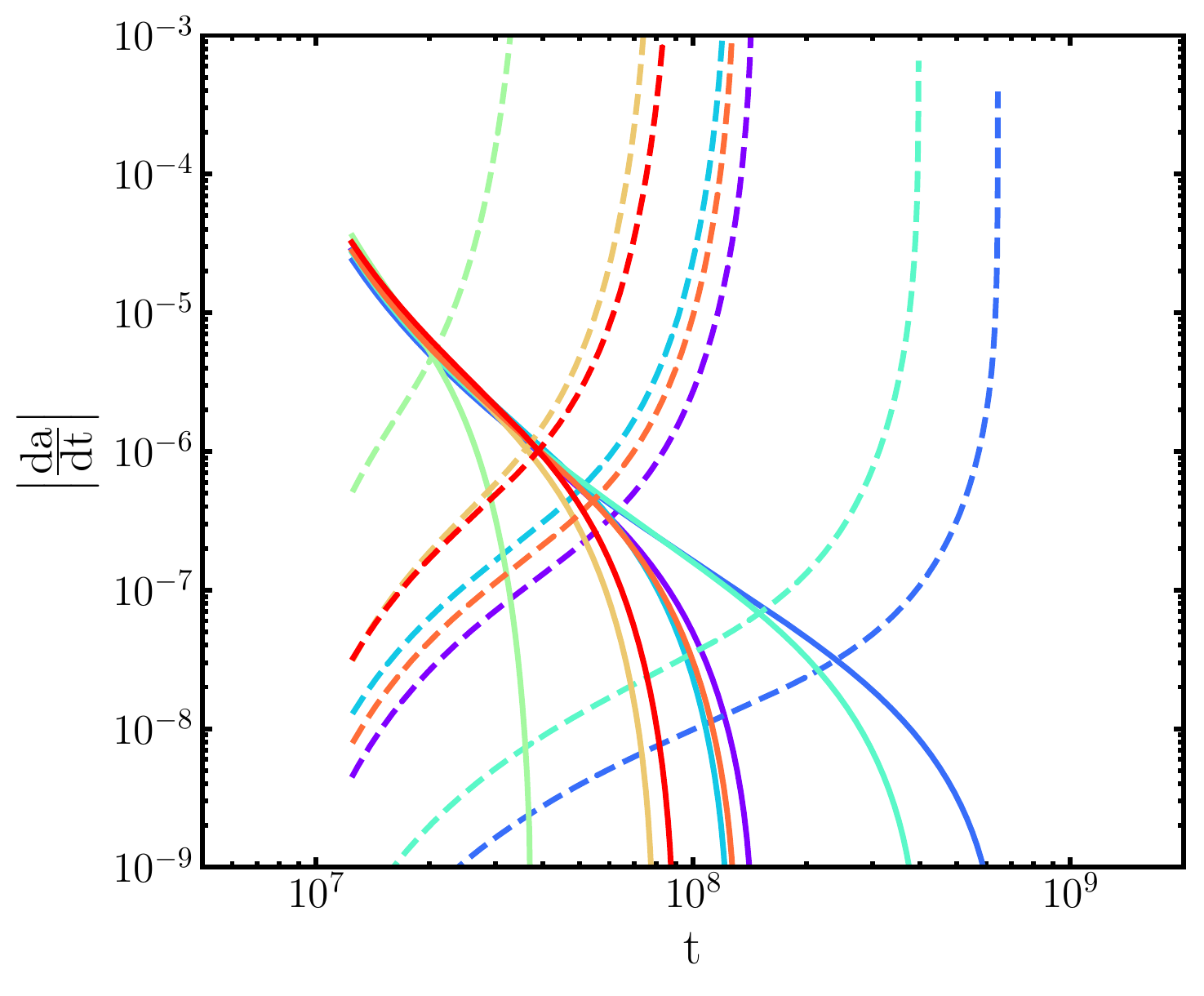}
    \caption{Hardening rate contributed by three-body encounters (solid lines) and GW emission (dashed lines) for all MR models, showing when GW emission begins to dominate the evolution of the binary. Scaling B is assumed. Colours are as described in Fig~\ref{fig:extrap_figure}.}
    \label{fig:hardening_frac_LR_scaleB}
\end{figure}
\begin{figure}
    \centering
    \includegraphics[width=\columnwidth]{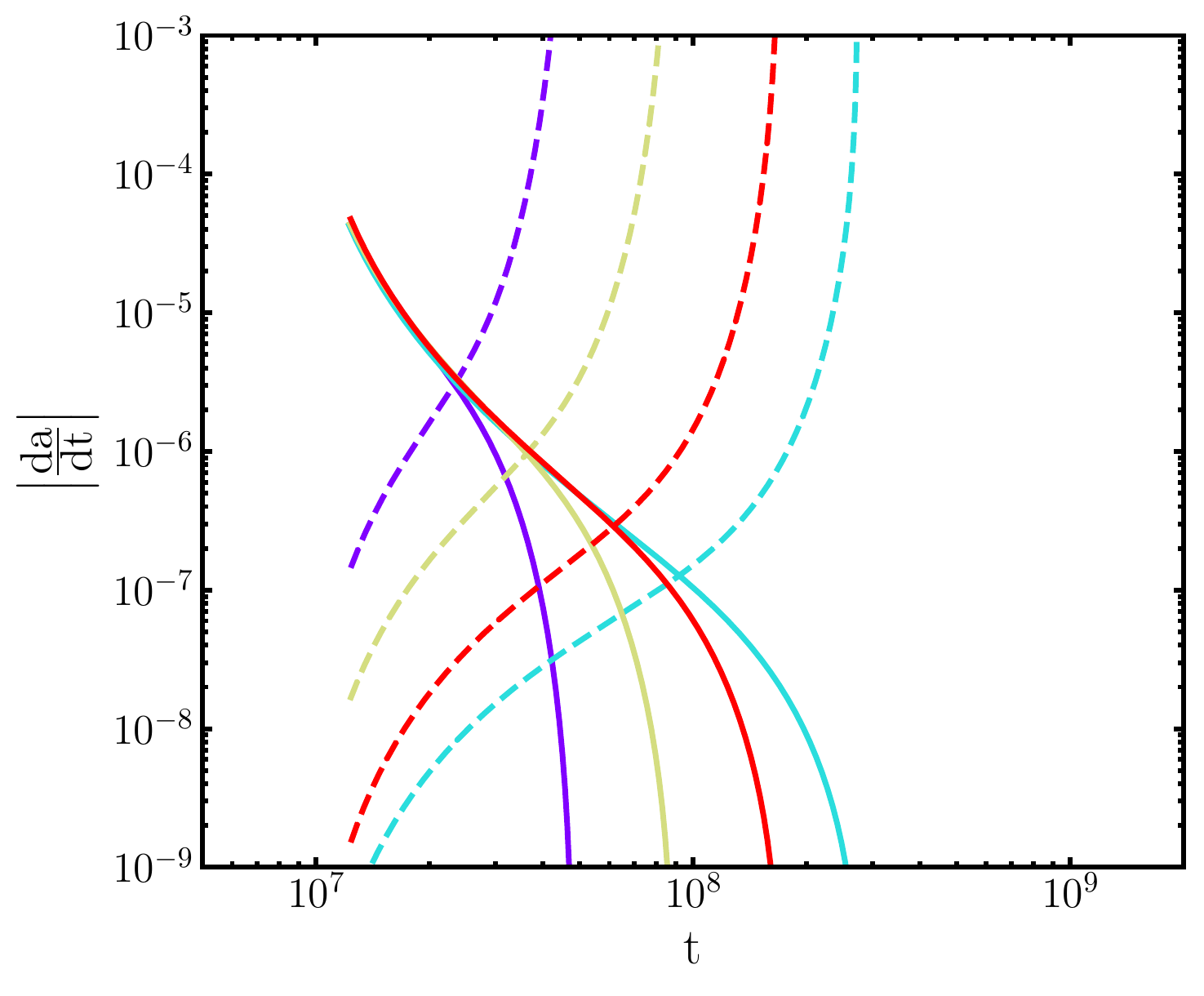}
    \caption{The same as Fig.~\ref{fig:hardening_frac_LR_scaleB} but for the HR models, with scaling B.}
    \label{fig:hardening_frac_HR_scaleB}
\end{figure}
This is confirmed in Figs.~\ref{fig:hardening_frac_LR_scaleB} and \ref{fig:hardening_frac_HR_scaleB} which show the hardening rate due to encounters (solid lines) and to GW emission (dashed lines). Once the rate of change of $1/a$ due to GW emission becomes comparable to that due to stellar interaction, the orbital evolution proceeds very quickly, shrinking and circularising the binaries until coalescence is reached.
Another interesting feature of Figs.~\ref{fig:hardening_frac_LR_scaleB} and \ref{fig:hardening_frac_HR_scaleB} is that GW emission is important for a significant fraction of time in evolving the binary to coalescence, due to the very large initial eccentricities. We note that in such cases a non-zero residual eccentricity may be present at coalescence, and this may affect the waveforms of LISA and PTA sources.

\section{Stochastic binary evolution}

\subsection{Eccentricity at binary formation}
Our simulations show, for the first time, significant stochasticity in the eccentricity of the binary at the time it becomes bound. An in-depth investigation of this phenomenon and the dependence on galactic and orbital parameters will be presented in a forthcoming work.  We attribute the stochasticity to the dependence of the binary's angular momentum evolution on stellar masses. This is supported by the observation that stochasticity is more significant in the MR and LR runs, where stellar masses are larger compared to the BHB mass. 

\begin{figure}
    \centering
    \includegraphics[width=\columnwidth]{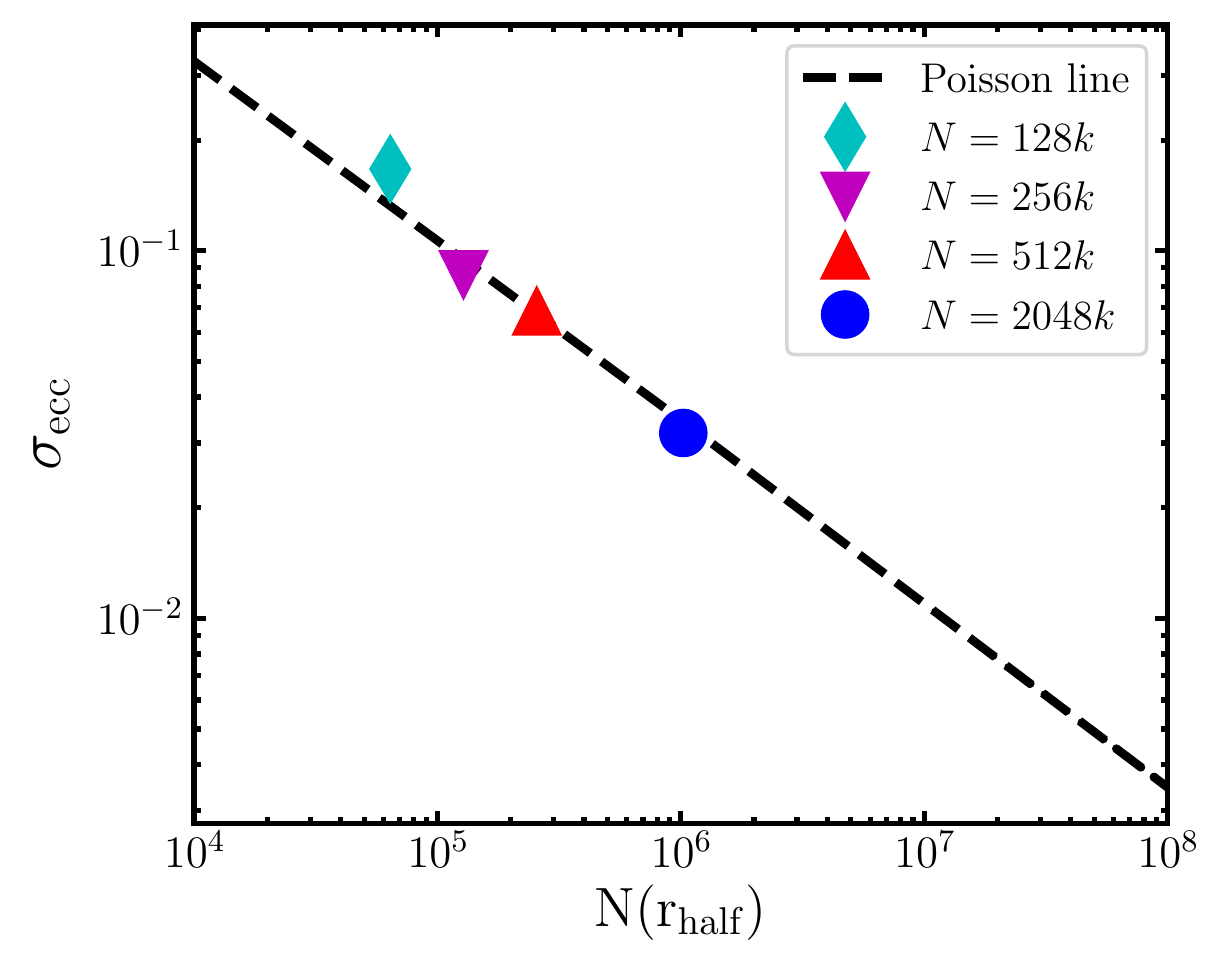}
    \caption{Dispersion of the eccentricity at binary formation as a function of the number of particles within the half mass radius $N(r_{\rm half})$ for the different resolutions. The  dashed line represents the Poisson standard error normalised to the $N=512$k resolution. The figure is extended to large $N$ values to allow for extrapolations. PR (diamond), LR (upside down triangle), MR (triangles) and HR (circle). }
    \label{fig:ecc_disp}
\end{figure}

Before establishing the effect of stochasticity on merger timescales, we compute the dispersion in the eccentricity at binary formation, $\se$, for all simulations. The dispersion is computed over a finite time interval $t = 150-160$ to reduce noise. The results are shown in Fig.~\ref{fig:ecc_disp} as a function of the number of stars within the half-mass radius of the resultant galaxy $N(r_{\rm{half}})$, showing a smaller dispersion in the HR runs. 
The figure also shows the predicted scaling with particle number for a Poisson random process, $\sigma \propto 1/\sqrt{N}$, normalised to the numerical value obtained for the MR models. The prediction matches well with the numerical dispersion measured for the HR runs, supporting the interpretation of the eccentricity spread as a stochastic process. The relation between $\se$ and the number of particles within the half mass radius can be written as
\begin{equation}
\sigma_{\rm ecc} = \frac{k}{\sqrt{N\left(r_{\rm half}\right) + k^2}}
\label{eq:ecc_disp_relation}
\end{equation}
where $k$ is a constant that for our models takes the value $k=39.7$. We include a constant term in the denominator to ensure that the dispersion converges for $N\rightarrow 0$. This relation can be used to calculate the resolution required to obtain a specific dispersion in eccentricity.

\subsection{The merger timescale}  
The dispersion in merger timescales obtained from our extrapolation to the GW emission phase is given in Table~\ref{tab:merger_time_params}. This shows a clear correlation with $\sigma_{\mathrm{ecc}}$ and a dependence on the scaling. Because more massive BHBs evolve faster and have a shorter GW timescale, variations in eccentricity have a smaller effect. A similar result has been observed by \citet{khan2015} who find that more massive BHBs spend less time in the three-body scattering phase.

\begin{figure}
    \centering
    \includegraphics[width=\columnwidth]{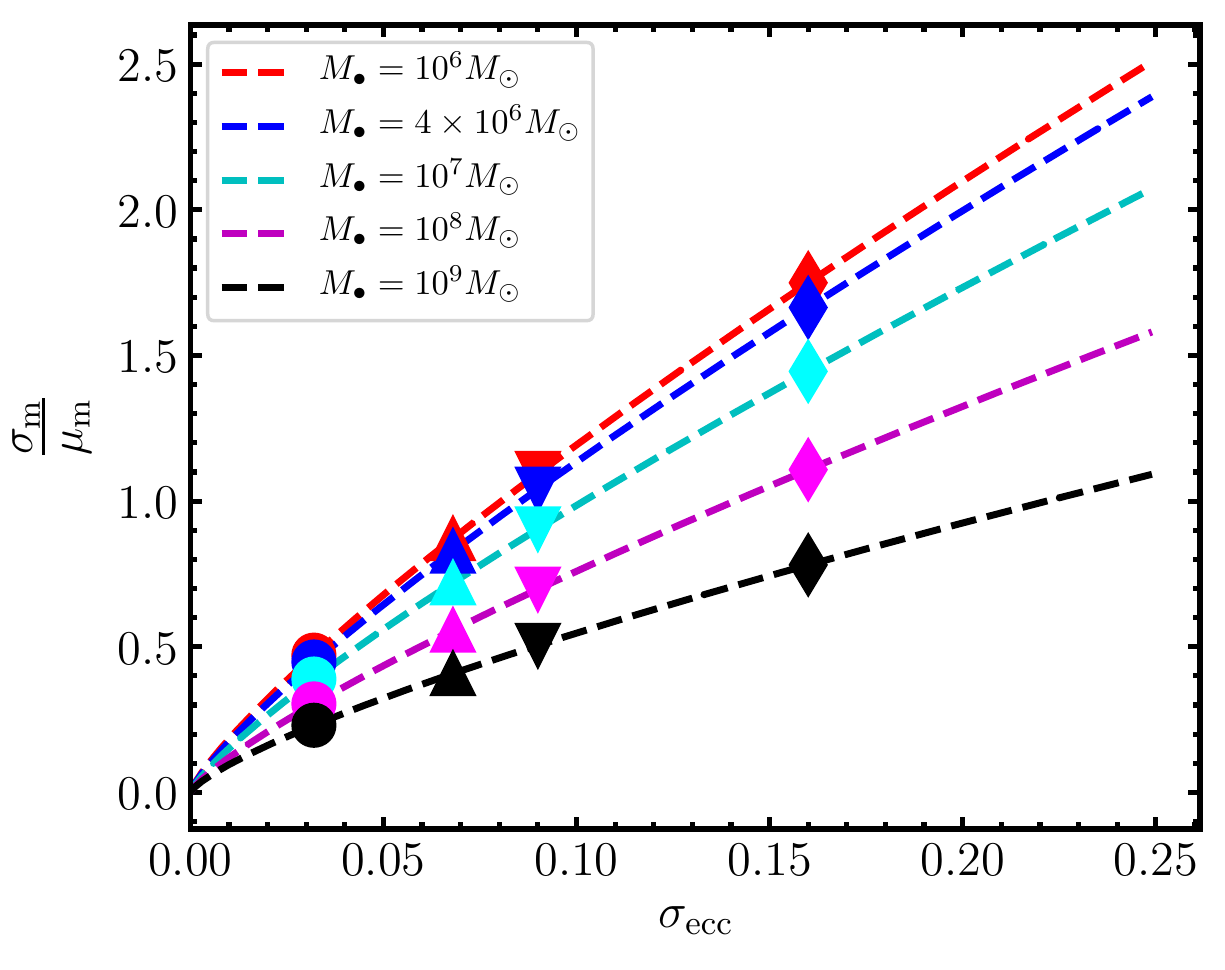}
    \caption{The \textit{coefficient of variation} of the merger time scale as a function of the dispersion in the eccentricity at binary formation for the five scalings considered, with SMBH mass increasing from top to bottom. PR (diamond), LR (upside down triangle), MR (triangles) and HR (circle).}
    \label{fig:merge_disp}
\end{figure}

We compute the \textit{coefficient of variation} for the merger timescale, i.e. the ratio of the dispersion to the mean, which is a dimensionless standardised measure of the dispersion. We then fit a power-law relation of the type
\begin{equation}
\label{eq:sigmam}
\frac{\sigma_{m}}{\mu_{m}} = C_{s}^{} \se^{n},   
\end{equation}
where $\mu_{m}$ is the average merger timescale and $C_{s}$ is a constant containing the scaling dependence. The results for the five scalings given in Table~\ref{tab:scaling} are shown in Fig.~\ref{fig:merge_disp}.
These show a clear dependence on scaling, with lower SMBH mass scalings having a larger dispersion in the merger timescale. Equation~\eqref{eq:sigmam} can be used to quantify the uncertainty in the merger timescale knowing the dispersion in eccentricity.

\begin{figure}
    \centering
    \includegraphics[width=\columnwidth]{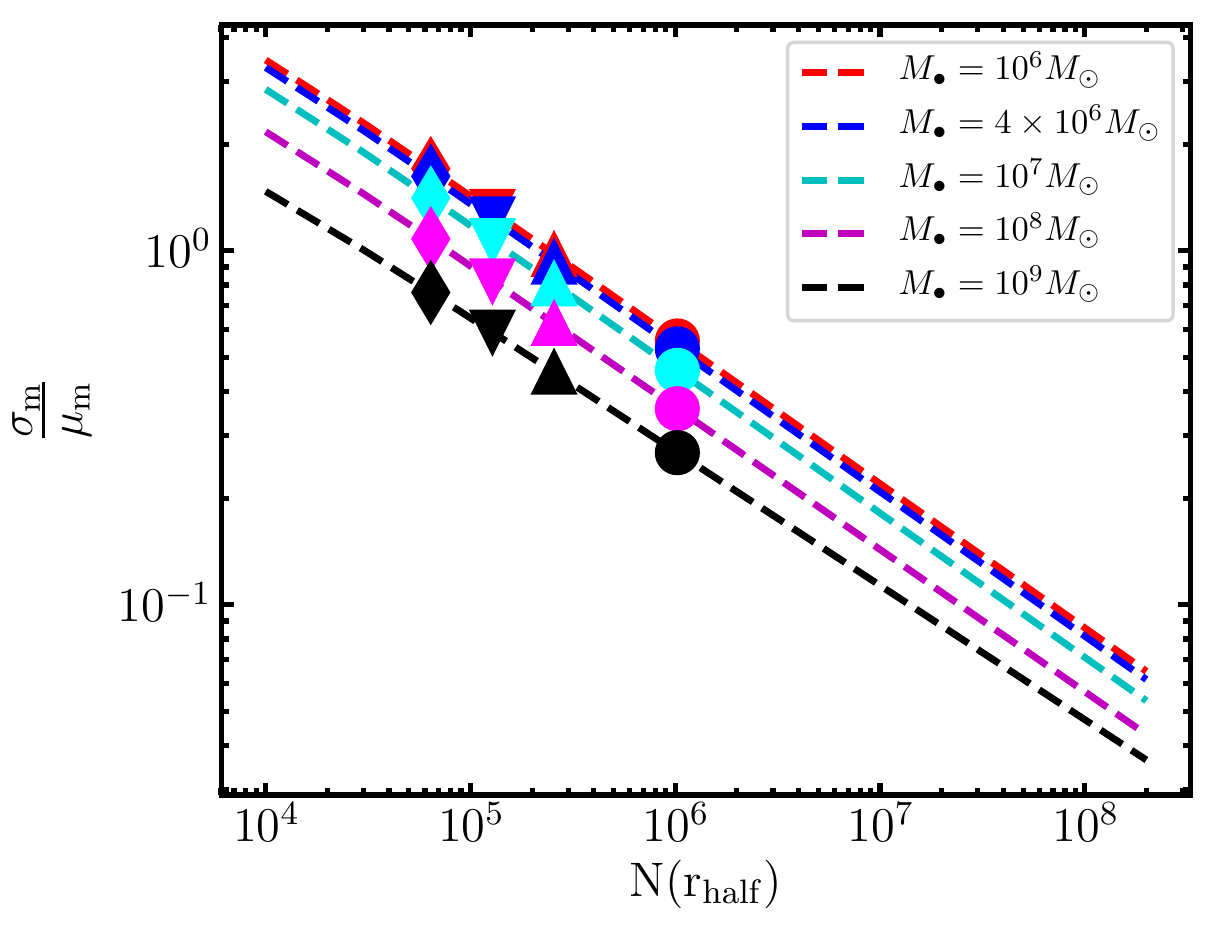}
    \caption{The \textit{coefficient of variation} of the merger timescale as a function of the number of particles within the half mass radius for all five scalings, with SMBH mass increasing from top to bottom. PR (diamond), LR (upside down triangle), MR (triangles) and HR (circle).}
    \label{fig:merge_disp_res}
\end{figure}

Combining this relation with equation~\eqref{eq:ecc_disp_relation}, we derive an expression for the coefficient of variation as a function of the number of particles within the half mass radius $N(r_{\rm half})$
\begin{equation}
\frac{\sigma_{m}}{\mu_{m}} = C_{s} k^n \left(N\left(r_{\rm half}\right) + k^2\right)^{-\frac{n}{2}}.
\label{eq:merge_disp_res_relation}
\end{equation}
The fitting parameters are given in Table~\ref{tab:merger_time_params}.
\begin{table}
\centering
\caption{Fitting parameters for the merger timescale relation as well as the number of particles within the half mass radius $N(r_{\rm {half}})$ to estimate the merger time to a $10\%$ uncertainty, measured in dex.}
\begin{tabular}{cccccc}
\hline 
Scaling & $\msbh$ & $C_{s}$ & $n$ & $k$ & $\frac{\sigma_{m}}{\mu_{m}} = 0.1$ \\
\hline 
A  & $10^{6}\msun$ & $7.79$ & $0.815$ & $39.7$ & $7.84$  \\
B  & $4 \times {10^{6}}\msun$ & $7.43$ &  $0.816$ & $39.7$ & $7.78$ \\
C  & $10^{7}\msun$ & $6.44$ & $0.815$ & $39.7$ & $7.64$  \\
D  & $10^{8}\msun$ & $4.83$ & $0.804$ & $39.7$ & $7.39$ \\
E  & $10^{9}\msun$ & $3.13$ & $0.758$ & $39.7$ & $7.14$ \\
\hline
\end{tabular}
\label{table:dispersion_relations}
\end{table}
The resulting relation with $N(r_{\rm half})$ is shown in Fig.~\ref{fig:merge_disp_res} for all five scalings.
We note that all the dependence on scaling is contained in the constant $C_s$, while $k$ and $n$ contain information on the galactic models and the orbital parameters of the merger.

By inverting equation~\eqref{eq:merge_disp_res_relation}, we are able to calculate the required resolution in order to obtain the merger timescale of the binary to a given accuracy. For LISA sources in the mass range $M_{\bullet}=10^6-10^7M_{\odot}$ the required resolution in order to accurately obtain the merger timescale to $10\%$ accuracy is well approximated by $N\left(r_{\rm half}\right) = C_{s}^{2.5}\left(4.4\times{10^{5}}\right)$,
where the scaling constant $C_{s}$ depends on the binary mass. From the scaling constants derived in Table~\ref{table:dispersion_relations}, we find that in order to accurately estimate the merger time-scale of a LISA source to a $10\%$ uncertainty, the resolution required within the half mass radius is in excess of ten million particles.
 
\section{Discussion and Conclusions}

We computed the merger timescale of black hole binaries formed in mergers of equal mass galaxies hosting central supermassive black holes. The evolution of the binaries is followed from the onset of the galactic merger through the hardening phase and to a separation smaller than the hard-binary separation. The FMM code \textsc{griffin}, which follows the BHB with direct summation, produces results consistent with the \textsc{$\phi$-grape} code. We considered different random realisations of the same model at different resolutions, up to two million particles. We found that, for the models considered here (a shallow cusp and large orbital eccentricity for the galaxy mergers), the eccentricity with which BHBs bind is highly stochastic, showing a dependence on the masses of the stars undergoing encounters with the binary. We verify that this spread in eccentricity decreases with particle number as a Poisson process, confirming its stochastic origin. The same process is responsible for Brownian motion of the binary.

We adopted a semi-analytic model to extrapolate the evolution of the binary's orbital elements beyond the $N$-body integrations, allowing us to determine the merger timescale due to emission of gravitational waves. The model adopted a fit of the time-dependent hardening rate over the whole $N$-body integration after binary pairing to estimate the change in orbital elements due to encounters with intersecting stars, as well as the classical \citet{peters1964} description of the evolution during the GW emission phase. We found a strong dependence of the merger timescale on the eccentricity dispersion.

We provided simple relations between the dispersion in the merger timescale and the dispersion in the eccentricity at binary formation, as well as the number of particles enclosed within the system's half mass radius. We estimated that particle numbers in excess of ten million are required to achieve a dispersion in the merger timescale of order $10\%$ of the mean value. Such particle numbers are currently beyond the capabilities of direct summation codes like \textsc{$\phi$-grape} and more efficient methods are required, for example the \textsc{griffin} code.

We note that the models chosen for this study are characterised by a shallow $\gamma=0.5$ inner cusp profiles and a large $e=0.9$ orbital eccentricity, a configuration most susceptible to perturbations and therefore to stochasticity. The large orbital eccentricity, in particular, which was chosen to mimic conditions found in cosmological simulations as well as to reduce computational time, may lead to BHBs flipping their orbital plane, becoming counter-rotating with respect to the overall stellar distribution. This will lead to a larger eccentricity at pairing and more significant perturbations \citep{khan2019}. We will present simulations of BHBs with steeper density profiles and less radial orbits in a forthcoming work.

 The chosen parameters are also expected to produce the most eccentric binaries, and in fact all BHBs formed in our simulations have  eccentricities larger than 0.9 at the time when GW emission becomes significant in the binary's evolution, and one binary reaches $e>0.99$.
This is a result of the three-body encounters driving the binary hardening, as already reported in several works \citep[e.g.][]{MMS07, sesana2010, sgd2011}. Large eccentricities are of consequence as they imply a faster coalescence as well as the possibility of detecting a residual eccentricity in the instrument's detection band \citep{PS2010}.

\section*{Acknowledgements}
We thank Elisa Bortolas, Fazeel Khan and Alberto Sesana for interesting discussions on the evolution of massive black hole binaries. The authors acknowledge the use of the Surrey Eureka supercomputer facility and associated support services. FA acknowledges support from a Rutherford fellowship (ST/P00492X/1) from the Science and Technology Facilities Council. MD acknowledges support by ERC-Syg 810218 WHOLE SUN.

\bibliographystyle{mnras}
\bibliography{bh_bib}

\bsp	
\label{lastpage}
\end{document}